\documentclass[floatfix,onecolumn,aps,prl,showpacs,amsmath,amssymb]{revtex4-1}
\usepackage[latin1]{inputenc}
\usepackage[dvips]{graphicx}

\usepackage{dcolumn}
\usepackage{bm}
\usepackage{dsfont}
\usepackage{color}
\usepackage{url}
\usepackage[colorlinks=true ,citecolor=blue]{hyperref}
\usepackage{amsmath,amssymb,esint}
\usepackage[table]{xcolor}

\definecolor{Nathanblue}{rgb}{0.,0.24,0.51}

\newcommand{\be}{\begin{equation}}
	\newcommand{\ee}{\end{equation}}
\newcommand{\bq}{\begin{eqnarray}}
	\newcommand{\eq}{\end{eqnarray}}

\begin{document}

\title{Topological Phase Transitions with Zero Indirect Band Gap}

\author{Giandomenico Palumbo}
\affiliation{School of Theoretical Physics, Dublin Institute for Advanced Studies, 10 Burlington Road,
	Dublin 4, Ireland}

\date{\today}

\email{giandomenico.palumbo@gmail.com}

\begin{abstract}
	\noindent Topological phase transitions in band models are usually associated to the gap closing between the highest valance band and the lowest conduction band, which can give rise to different types of nodal structures, such as Dirac/Weyl points, lines and surfaces. In this work, we show the existence of a different kind of topological phase transitions in one-dimensional systems, which are instead characterized by the presence of a robust zero indirect gap, which occurs when the top of the valence band coincides with the bottom of the conduction band in energy but not in momentum. More specifically, we consider an one-dimensional model on a diamond-like chain that is protected by both particle-hole and chiral-inversion symmetries. At the critical point, the system supports a Dirac-like point. After introducing a deforming parameter that breaks both inversion and chiral symmetries but preserves their combination, we observe the emergence of a zero indirect band gap, which results to be related to the \emph{persymmetry} of our Hamiltonian. Importantly, the zero indirect gap holds for a range of values of the deforming parameter. We finally discuss the implementation of the deforming parameter in our tight-binding model through time-periodic (Floquet) driving.
\end{abstract}

\maketitle

\section*{Introduction}

Topological phase transitions play an important role in deepening our understanding of topology in quantum matter.
In band systems, the closing of the direct gap is associated to a plethora of topological critical phases, which can give rise to different kinds of topologically protected momentum-space defects ranging from Dirac and Weyl points to Dirac and Weyl nodal lines and surfaces \cite{Armitage,Rappe}.
However, critical phases with a zero-indirect band gap are much less understood. They can exist when the top of the valence band coincides with the bottom of the conduction band in energy but not in momentum. In other words, differently from more common topological semimetals, the closing of a gap here does not imply the existence of gapless nodal structures. In two dimensions, a robust zero indirect gap without fine-tuning has been shown to emerge in some Chern phases on the Lieb lattice \cite{Palumbo}. Its robustness is supported by the existence of a range of values of certain parameters of the model that keep at zero the indirect gap. This represents a novel scenario with respect to other quantum systems where a zero indirect gap ha been shown to exist only for fine-tuned values of certain physical parameters \cite{Teppe,Irie,Mohan}. In this work, we provide the first example of a topological phase transition with a robust zero indirect band gap in one dimension by considering a model with three species of spinless fermions on a diamond-like chain. Differently from previous works concerning diamond and trimer chains \cite{Vidal, Bercioux, Bercioux2, Goldman2, Alvarez, Bercioux3}, our Hamiltonian results to be characterized by an overlooked symmetry in topological matter, known in mathematical literature as persymmetry \cite{Huckle}. A persymmetric matrix is a square matrix which is symmetric with respect to the anti-diagonal.  Firstly, we will show that our model supports both particle-hole and chiral-inversion symmetries with a critical phase transition characterized by the emergence of a three-fold degenerate Dirac-like point. Secondly, by adding a deforming parameter that preserves both persymmetry, particle-hole and chiral-inversion symmetries but breaks independently inversion and chiral, we will show that the Dirac-like point splits giving rise to three non-degenerate bands where there appears a zero indirect gap between the lowest and middle bands. Thus, at filling 1/3 the system remains semimetallic even in absence of band touching points.
Importantly, the zero indirect band gap persists for different values of the particle-hole-symmetric deforming parameter $m$ and survives even in presence of further terms that break particle-hole and chiral-inversion symmetries but preserve persymmetry. In fact, persymmetry represents the crucial feature of our novel gapless phase. We will show that once reached a critical value $m=m^{*}$, the valence and conduction bands touch each other and for $m>m^{*}$ an indirect gap opens giving rise to an insulating phase even at filling 1/3.
Finally, we will show that $m$ can be naturally induced in our system through Floquet engineering, namely via time-periodic driving, which can be implemented in a suitable one-dimensional synthetic-matter setup such as ultracold atoms.

\section*{One-dimensional system}
Here, we start introducing a one-dimensional system made by three fermion species on a diamond-like chain as shown in Fig.1.
\begin{figure}[htp]
	\begin{center}
		\includegraphics[scale=0.5]{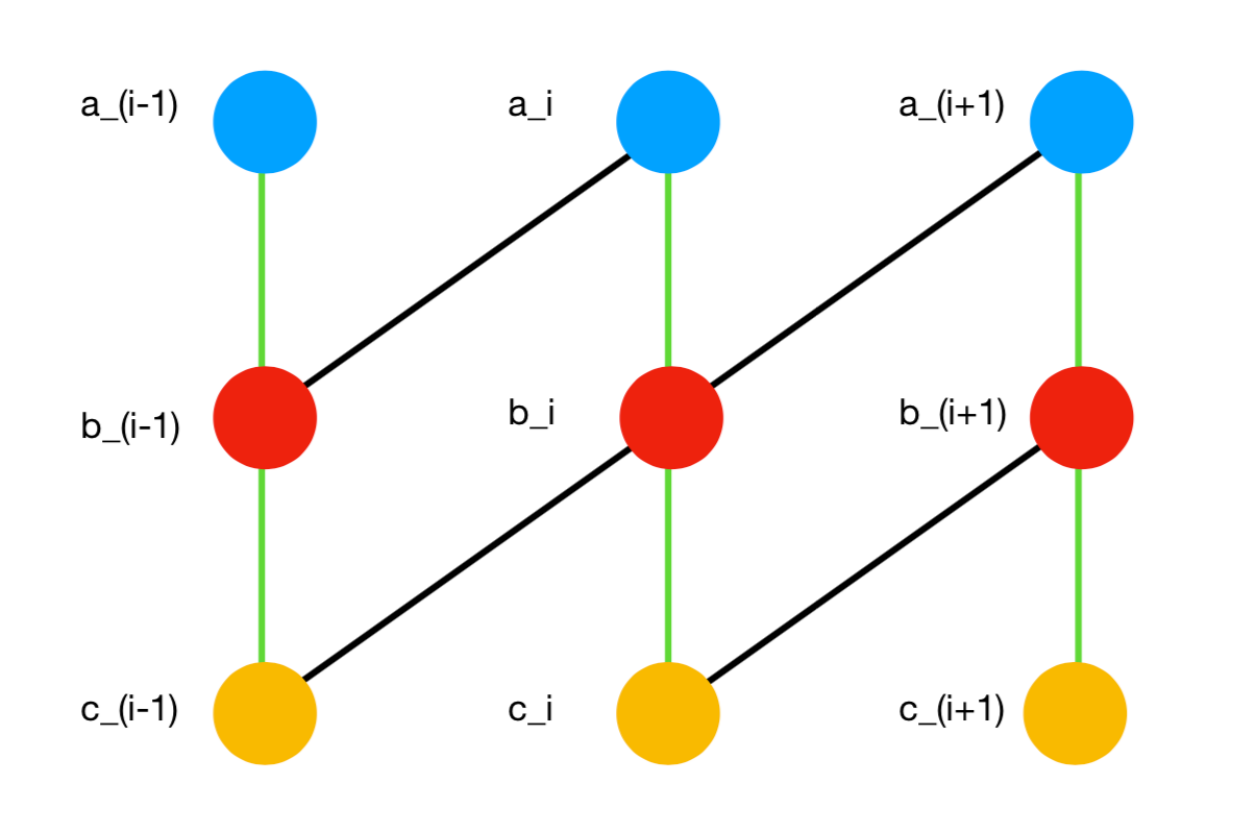}
			\caption{Diamond-like chain with three species of spinless fermions represented by blue (a), red (b) and yellow (c) dots. Green lines represent on-site inter-species hopping while dark lines are related to nearest-neighbour inter-species hopping.}
	\end{center}
\end{figure}
The corresponding real-space tight-binding Hamiltonian is given by
\begin{eqnarray}
H_{rm}=\sum_i \left[J (a^{\dagger}_{i+1}b_i+b^{\dagger}_{i+1}c_i)-h (a^{\dagger}_i b_i-b^{\dagger}_i c_i)+ i m\, a^{\dagger}_i c_i + h.c.\right],
\end{eqnarray}
where all the physical parameters $J$, $h$ and $m$ are taken real (here, we have taken a collapsed bases, where all the three fermions live on the same given site). Crucially, the deforming parameter $m$ plays a central role for the appearance of a zero indirect band gap as we will show below.
The corresponding momentum-space Hamiltonian is then given by
\begin{eqnarray}
H_p = \psi^\dagger H \psi,
\end{eqnarray}
where $\psi=(a_p, b_p, c_p)^T$ and
\begin{eqnarray}
H= \left( \begin{array}{ccc}
0 & J e^{-i p}-h & i m \\
J e^{i p}-h & 0 & J e^{-i p}-h \\
-i m & J e^{i p}-h & 0 \end{array} \right).
\end{eqnarray}
This Hamiltonian does not have an analytic spectrum, i.e. the three real eigenvalues have only numerical expressions for $m \neq 0$. For $m=0$ the system has a perfectly flat middle band and all the three bands touch each other at a single Dirac-like point when $J=h$. For $J>h$ the system supports a gapped topological phase characterized by a quantized winding number. In fact, this topological invariant is protected by
particle-hole symmetry, i.e.
\begin{eqnarray}
U_{PH} H (p) U_{PH}^{-1} = -  H^*(-p) ,
\end{eqnarray}
where
\begin{eqnarray}
	U_{PH} = \left( \begin{array}{ccc}
		1 & 0 &0 \\
		0 & -1 & 0 \\
		0 & 0 & 1 \end{array} \right),
\end{eqnarray}
while time-reversal symmetry is absent for $m\neq 0$. In the case $J<h$, the corresponding gapped phase is topologically trivial. Because $m$ does not break particle-hole symmetry and for small values it does not induce any phase transition, we can set it equal to zero for simplicity in order to analytically calculate the winding number.
The Abelian Berry connection for the lowest band is given by
\begin{eqnarray}
	A_{p}= <\psi |i \partial_{p}| \psi > = \frac{1}{2}\left(1+\frac{J}{J-h\,e^{i p}}-\frac{h}{h-J\, e^{i p}}\right),
\end{eqnarray}
where $| \psi>$ is the momentum-space Bloch eigenvector related to the lowest band. The corresponding winding number $w$ can be then calculated by integrating $A_{p}$ on the first Brillouin zone, i.e.
\begin{eqnarray}
	w= \int_{-\pi}^{\pi} dp A_{p} = 2 \pi,
\end{eqnarray}
for $J>h$ and $w=0$ otherwise.
 This result is compatible with the fact that one-dimensional topological insulators protected by particle-hole symmetry are in class D and are characterized by a $Z_2$ topological invariant \cite{Ryu}.
 Importantly, our model also supports a further symmetry given by
\begin{eqnarray}
U_{CI}	H (p) U_{CI}^{-1}= -  H (-p),
\end{eqnarray}
where
\begin{eqnarray}
	U_{CI} = \left( \begin{array}{ccc}
		0 & 0 &1 \\
		0 & -1 & 0 \\
		1 & 0 & 0 \end{array} \right),
\end{eqnarray}
which is known as chiral-inversion symmetry \cite{Jin}. Although the deforming parameter $m$ breaks both chiral and inversion symmetries, their combination, i.e. chiral-inversion is always preserved. Although in class D, our system cannot be identified with a topological superconductor. However, a doubled version of the above model could give rise to a proper topological superconducting phase due to the possible construction of a Bogoliubov-de Gennes Hamiltonian.
We now focus on the most peculiar feature of this model, namely the existence of a robust zero indirect gap.\\
At filling 1/3 (i.e. filled lowest band) and for $J=h$ and $|m| \leq \sqrt{2 J^2}$, the system behaves like a semimetal although the lowest and middle bands do not touch each other as shown in Fig.2b.
\begin{figure}[!h]
		\centering
		\includegraphics[width=.4\linewidth]{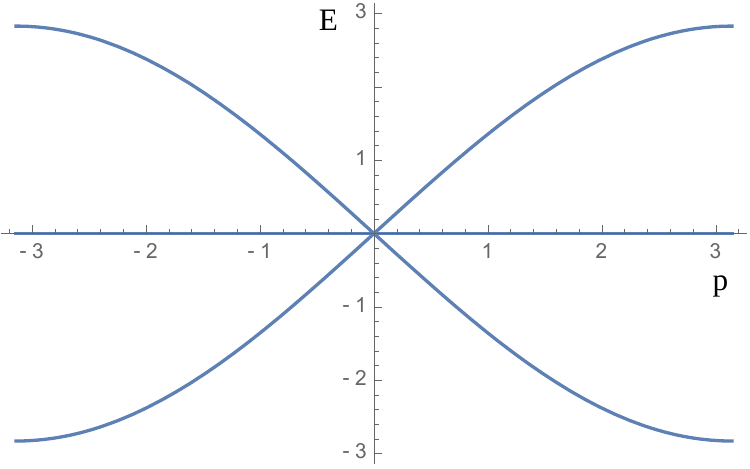}
		\includegraphics[width=.4\linewidth]{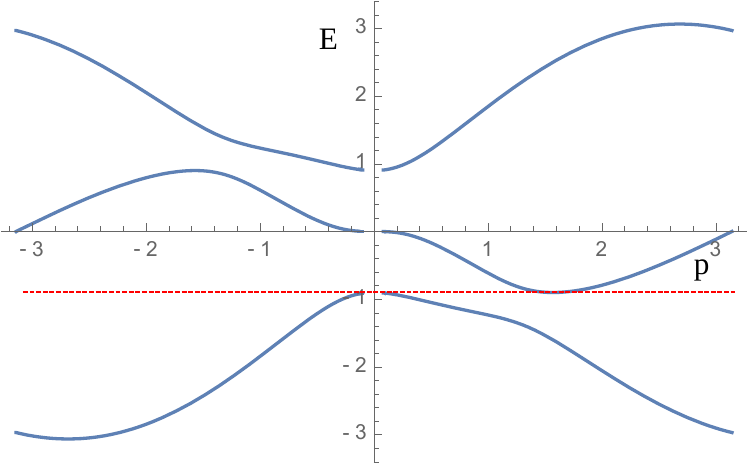}
		\caption{Left - $J=h=1$, $m=0$. The system is a conventional semimetal at half filling and supports a single Dirac-like point. Right - $J=h=1$, $m= 0.9$. The system supports a robust zero indirect band gap being semimetallic at filling 1/3. The chemical potential is represented by the dashed red line.}
			\includegraphics[width=.4\linewidth]{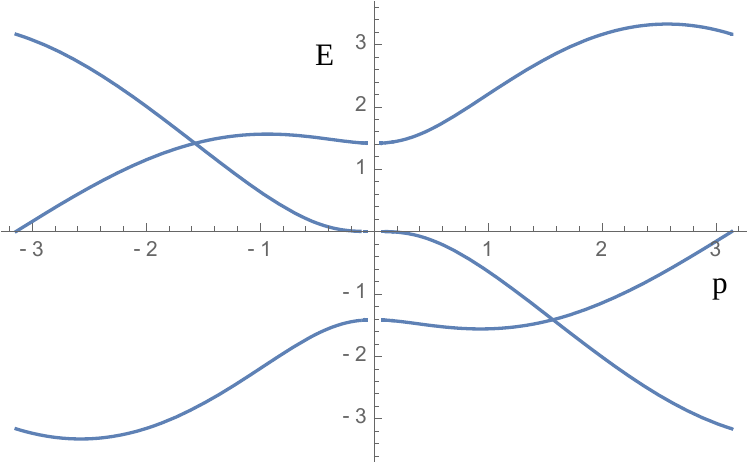}
				\includegraphics[width=.4\linewidth]{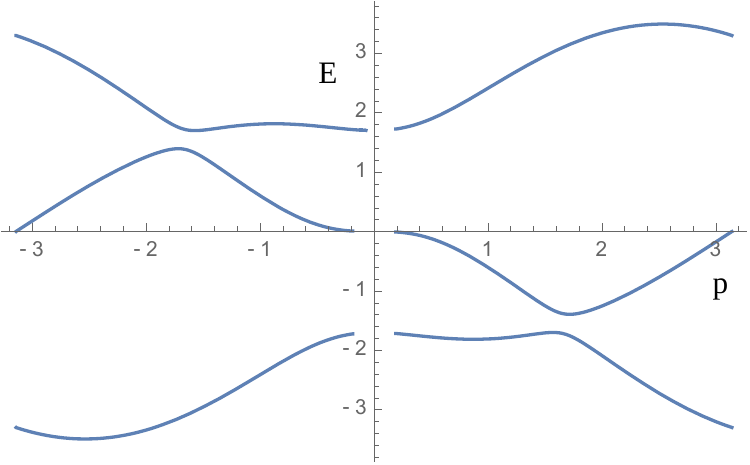}
		\caption{Left - $J=h=1$, $m= \sqrt{2}$. Here, we observe a non-trivial phase transition where the bands touch each others at some special Dirac points. Right - 
	$J=h=1$, $m=1.7$. The system has a finite direct gap.}
		\label{fig:sub2}
\end{figure}
In our knowledge, this model presents the first example of a one-dimensional semimetallic phase with a robust zero indirect band gap. Moreover, the system goes through a further phase transition at $|m|=\sqrt{2 J^2}$, i.e. when the valence and conduction bands touch each other via a Dirac point. For $|m| > \sqrt{2 J^2} $ the system becomes insulating with a finite direct gap.
Besides the symmetries mentioned above, $H$ is also a persymmetric matrix, namely it is symmetric with respect to the anti-diagonal. Although, persymmetry has been already studied in certain quantum systems \cite{Huckle}, in our knowledge, it has been overlooked in topological matter.
It plays a central role in our context because it is somehow related to the existence of the zero indirect band gap, which is not related to neither particle-hole nor chiral inversion symmetry. It is straightforward to show this by adding some on-site terms given by
\begin{eqnarray}
	H_f=\sum_i f \left(a^{\dagger}_{i}a_i-b^{\dagger}_{i}b_i+c^{\dagger}_i c_i\right),
\end{eqnarray}
with $f$ a real parameter. These diagonal terms break both particle-hole and chiral inversion symmetries but keep persymmetry.Similarly, we can also introduce the following interspecies hopping term
\begin{eqnarray}
	H_g=\sum_i g \left(a^{\dagger}_i c_i + h.c.\right),
\end{eqnarray}
with $g$ a real parameter such that only persymmetry is preserved. Thus, although both $H_{rf}=H_{rm}+H_f$ and $H_{rg}=H_{rm}+H_g$ are topologically trivial nevertheless they can still support a zero indirect band gap for certain values of the deforming parameters $m$, $f$ and $g$. This is a clear indication of the crucial importance of persymmetric Hamiltonians in our discussion.

\section*{Floquet theory}
In this section, we show that the deforming term $m$ can be naturally induced through Floquet engineering. Quantum systems under time-periodic driving can be in fact
described through the Floquet theory, which represents a temporal version of the Bloch theorem \cite{Kitagawa, Goldman, Eckardt}.
Among other things, it has been shown, for instance, that time-periodic driving can open a topological gap in two dimensions \cite{Oka} and induce nodal lines and surfaces in three-dimensional Dirac systems \cite{Salerno}.
In our case, we consider the following time-dependent Hamiltonian
\begin{eqnarray}
H_t=H_r + V(t),
\end{eqnarray}
where
\begin{eqnarray}
	H_r=\sum_i \left[J(a^{\dagger}_{i+1}b_i+b^{\dagger}_{i+1}c_i)-h(a^{\dagger}_i b_i+b^{\dagger}_i c_i)+ h.c.\right],
\end{eqnarray}
and
\begin{eqnarray}
	V(t)=G_1 \cos \omega t+ G_6 \sin \omega t,
\end{eqnarray}
where $\omega$ is the frequency of the drive and 
\begin{eqnarray}
G_1= \left( \begin{array}{ccc}
0 & 1 & 0 \\
1 & 0 & 0 \\
0 & 0 & 0 \end{array} \right), \hspace{0.3cm}
G_6= \left( \begin{array}{ccc}
0 & 0 & 0 \\
0 & 0 & 1 \\
0 & 1 & 0 \end{array} \right),
\end{eqnarray}
are nothing but two Gell-Mann matrices.
In the following, we assume that the drive energy $\hbar \omega$ is very large compared to all other energy scales in the system and we then set $\hbar\!=\!1$, except otherwise stated. 
The corresponding effective Floquet Hamiltonian is then defined as 
\begin{align}
	H_F=\frac{i}{T}\, \log \left[\mathcal{T} e^{-i \int_{0}^{T}dt\, H_{t}(t)}\right]\label{Floquet},
\end{align}
where $T=2 \pi/\omega$ is the period, while $\mathcal{T}$ represents the time-ordering.
The Floquet Hamiltonian can be systematically expanded in powers of $1/\omega$, employing standard procedures
\begin{eqnarray}
H_{F}=H_r+ \frac{i}{2 \omega}[G_1,G_6]+\frac{1}{4 \omega^2}([[G_1,H_r],G_1]+[[G_6,H_r],G_6])+...
\end{eqnarray}
Up to first order in $1/\omega$, we obtain the following identification
\begin{eqnarray}
	H_F \simeq H_{rm},
\end{eqnarray}
with
\begin{eqnarray}
m\equiv\frac{1}{2 \omega}.
\end{eqnarray}
At very high frequency, the first order is dominant and the effective Floquet Hamiltonian basically coincides with the static Hamiltonian $H_{rm}$. In any case, $H_{F}$ preserves particle-hole symmetry even in presence of the second-order correction.

\section*{Conclusions and outlook}
Summarizing, in this work we have proposed a new one-dimensional system that is characterized by persymmetry and a robust zero indirect band gap.
We have emphasized how Dirac-like points can slip by inducing a zero indirect gap and have shown that in our model the deforming parameter can be induced through a time-periodic driving. Although we have focused on a very specific example, these kinds of phase transitions can occur in any dimension as already shown in Ref.\cite{Palumbo} in a two-dimensional model. Moreover, time-reversal-symmetric models could naturally support zero-indirect-gap phases as we will show in future work. Another interesting point would be to figure out the possible existence of zero indirect gaps in suitable pseudo-Hermitian systems \cite{Zhu}.
Beyond the free-fermion case, it would be important to analyse the interacting regime of our system following several approaches and techniques already discussed in literature for other models \cite{Barbiero,Barbiero2}. A related question would concern the possible existence of a Luttinger liquid phase although in our case the zero indirect band gap would prevent us to study any linearized interacting phase around a Dirac point in the continuum limit.
We believe that our work deepens the current understanding of topological phase transitions. Moreover, synthetic matter setups, such as ultracold atoms, represent ideal platforms where to experimentally simulate and test our model.

\end{document}